\documentclass{article}
\usepackage[utf8]{inputenc}
\usepackage{graphicx}
\usepackage[a4paper=true]{hyperref}

\title{Status of the IAU Meteor Data Center}
\author{Regina Rudawska$^1$, M\'{a}ria Hajdukov\'{a}$^2$, Tadeusz J. Jopek$^3$,\\ Lubo\v{s} Neslu\v{s}an$^4$, Mari\'{a}n Jakub\'{i}k$^4$, J\'{a}n Svore\v{n}$^4$}
\date{\small{$^1$RHEA group/ESA ESTEC, The Netherlands, $^2$Astronomical Institute of the Slovak Academy of Sciences, Bratislava, Slovak Republic, $^3$Astronomical Observatory Institute, Faculty of Physics,  A.M. University, Poznań, Poland, $^4$Astronomical Institute of the Slovak Academy of Sciences, Tatranská Lomnica, Slovak Republic}}

\begin{document}
\maketitle
\noindent \small{Accepted: Proceeding of IMC2021, to be published in WGN, the Journal of the IMO.}

\subsection*{\centering Abstract}
Since 2007, the Meteor Data Center (MDC) has had two components: the ''Orbital database'' (OD) and the ''Shower database'' (SD).
The orbital part is in charge of the eﬀicient collection, checking, and dissemination of geocentric parameters and orbits of individual orbits. It also acts as a central depository for meteoroid orbits obtained by different techniques: photographic, television, video, CCD and radar. 
\\
The shower database collects the geocentric and orbital parameters of the meteor showers and meteoroid streams. It is not an archive  of all information related to meteor showers, its primary task is to give unique names and codes to new meteor showers (streams). The SD  acts in conjunction with the Working Group on Meteor Shower Nomenclature of International Astronomical Union (IAU) Commission F1, "Meteors, Meteorites, and Interplanetary Dust".
\\
In our paper, we give a concise description of the IAU MDC database, its origin, structure and, in particular, 
the current requirements for the introduction of new orbital and shower data. 

\section{Introduction}
The Meteor Data Centre (MDC)\footnote{\url{https://www.iaumeteordatacenter.org/}}
was established at the General Assembly of the IAU held in 1982, in Patras, and at that time compiled only orbital data. In 2006, in Prague, Commission 22 of the IAU established a Task Group for Meteor Shower Nomenclature, (see \cite{2007IAUTB..26..140S}).~\footnote{In 2009 the Task Group was transformed into the Working Group on Meteor Shower Nomenclature.} In the following year, the meteor shower database was created as part of the IAU MDC, and two years later, in 2009, in Rio de Janeiro, for the first time in the history of meteor astronomy, $64$ meteor showers were oﬀicially named by the GA IAU, (see \cite{2010IAUTB..27..158B,2010IAUTB..27..177W,2011msss.conf....7J}). The official names of the 
subsequent 48 showers were approved at the GA IAU held in Beijing and Honolulu (see \cite{2014me13.conf..353J,2017P&SS..143....3J}).

Over the years, the structure of the MDC database and the rules of its management have changed, 
see \cite{1987PAICz..67..201L,1994IAUS..160..497L,2003EM&P...93..249L,2011msss.conf..338P,2014EM&P..111..105N,2020P&SS..18204821J}.

\begin{table}[!ht]
\centering
\begin{tabular}{lr}
\hline
 {Catalogue/Source} &   {Number of orbits} \\
\hline
{Photographic/various data} & 4 873 \\
{Video/CAMS} &  110 521   \\
{Video/SonotaCo} & 353 231 \\
{Radar/Hissar} &  8 916 \\
\hline
\end{tabular}
\label{tab:orbits}
\caption{IAU MDC orbital data status October 2021.}
\end{table}

In this work, the words meteoroid stream and meteor shower can be understood as synonyms. Their current definition adopted by the IAU is available in the work of \cite{2017JIMO...45...91K}.

\begin{figure*}
\centerline{\includegraphics[width = 5 cm, angle = -90]{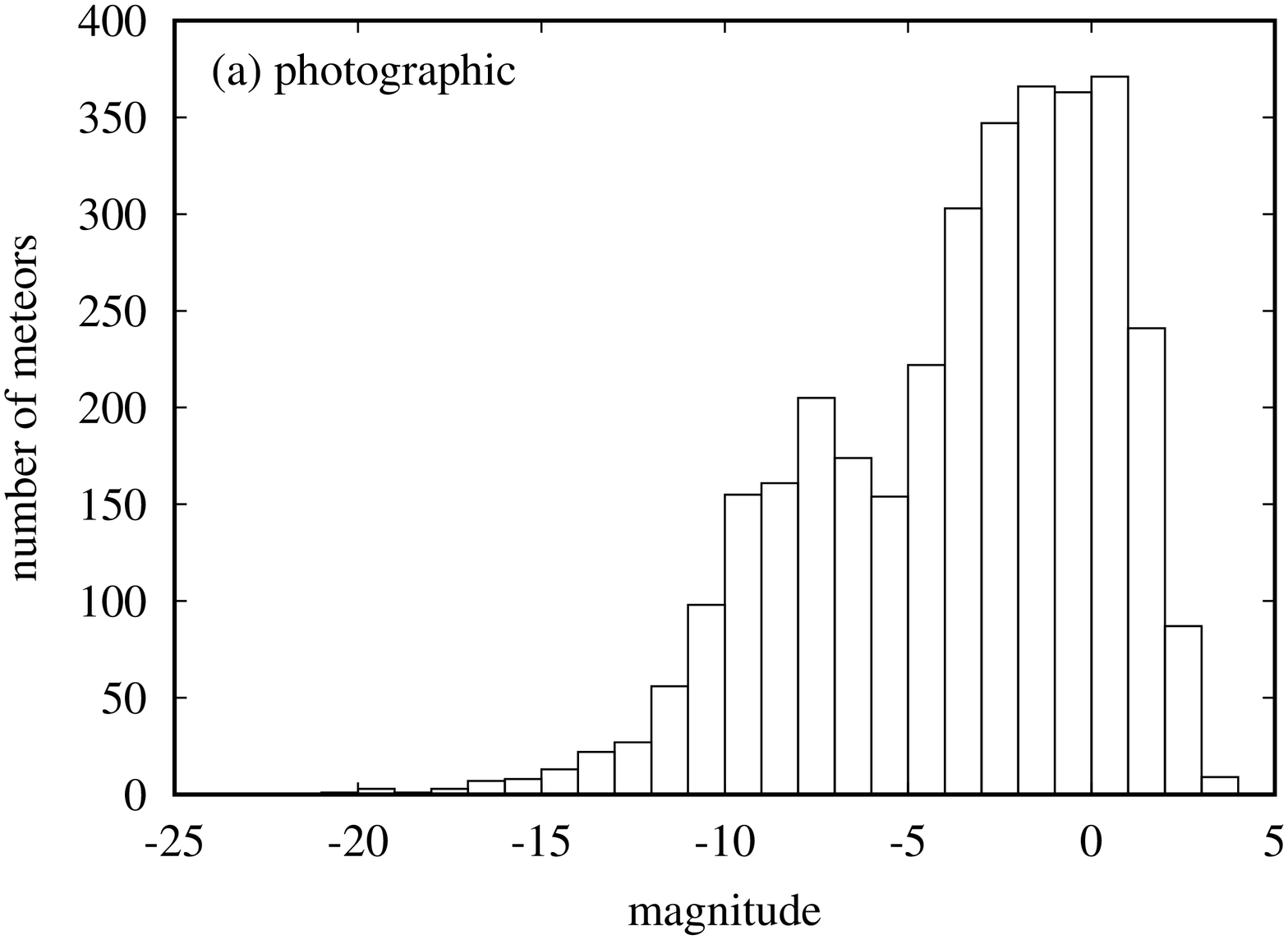}
  \includegraphics[width = 5 cm, angle = -90]{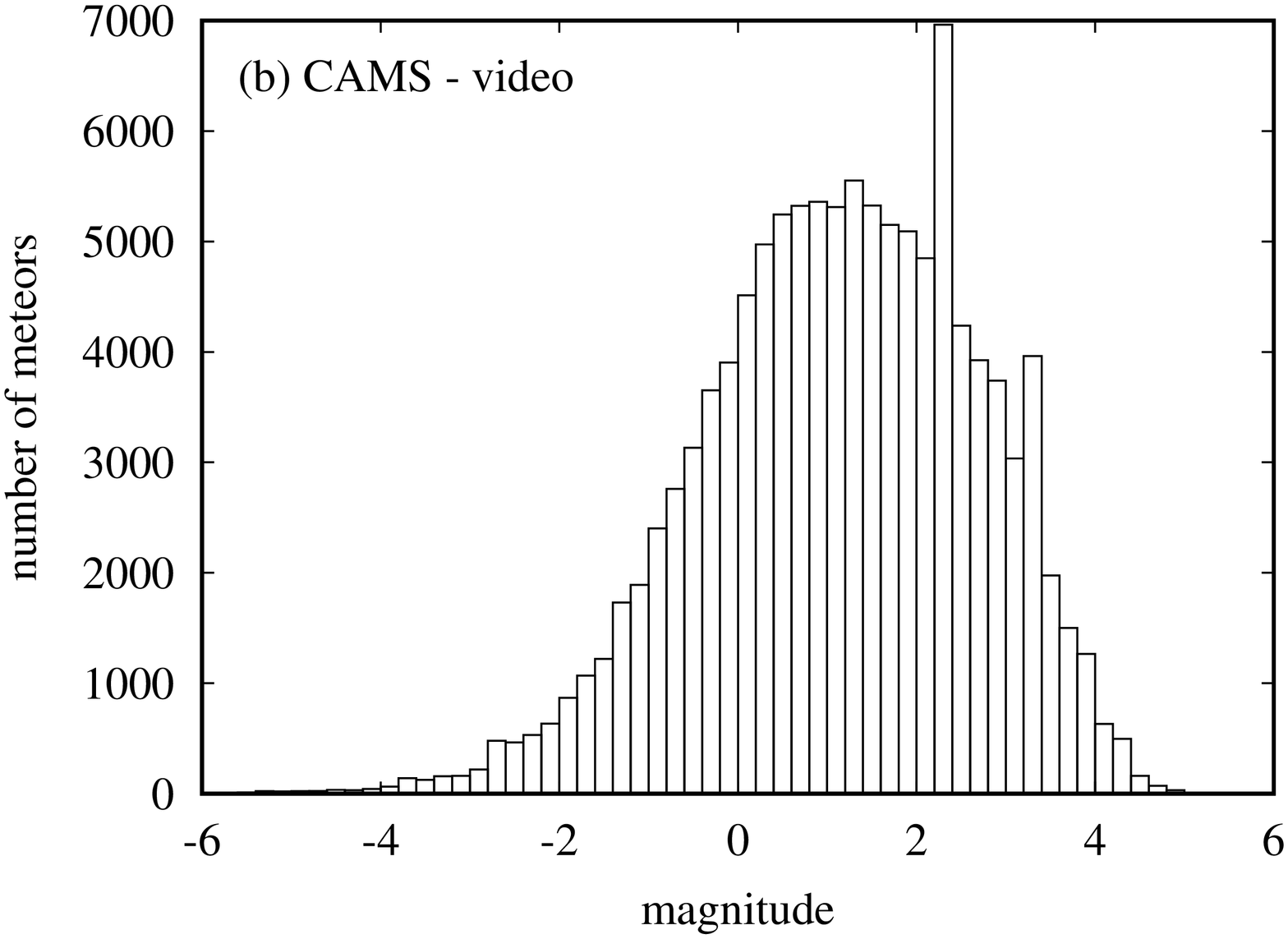}}
\caption{The magnitude distribution of all photographic (panel a) and CAMS video (b) meteors. Eleven video meteors brighter than $-6^{m}$ are out of the shown range.}
\label{fig:histF}
\end{figure*}

\begin{figure*}
\centerline{\includegraphics[width = 5 cm, angle = -90]{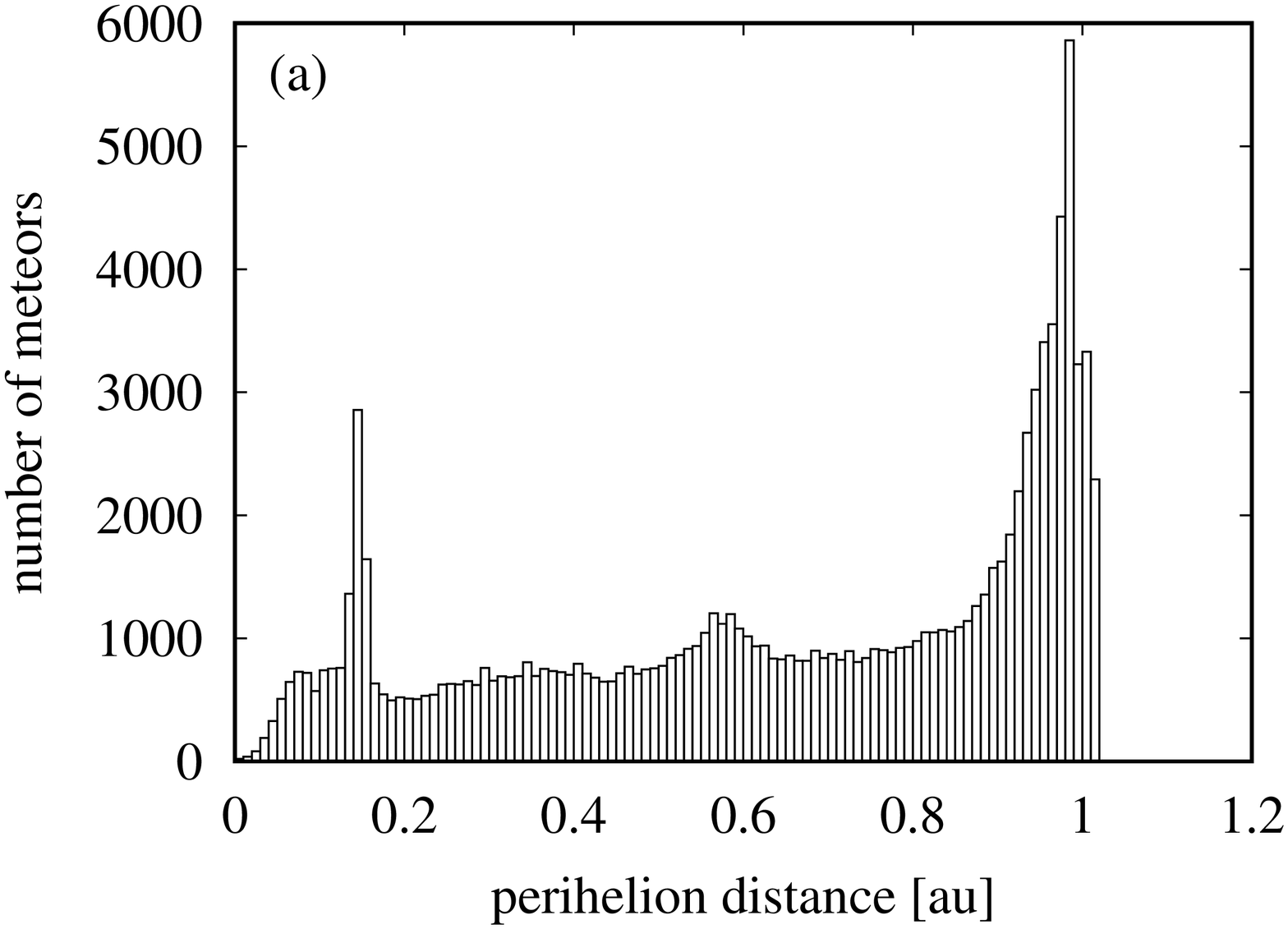}
  \includegraphics[width = 5 cm, angle = -90]{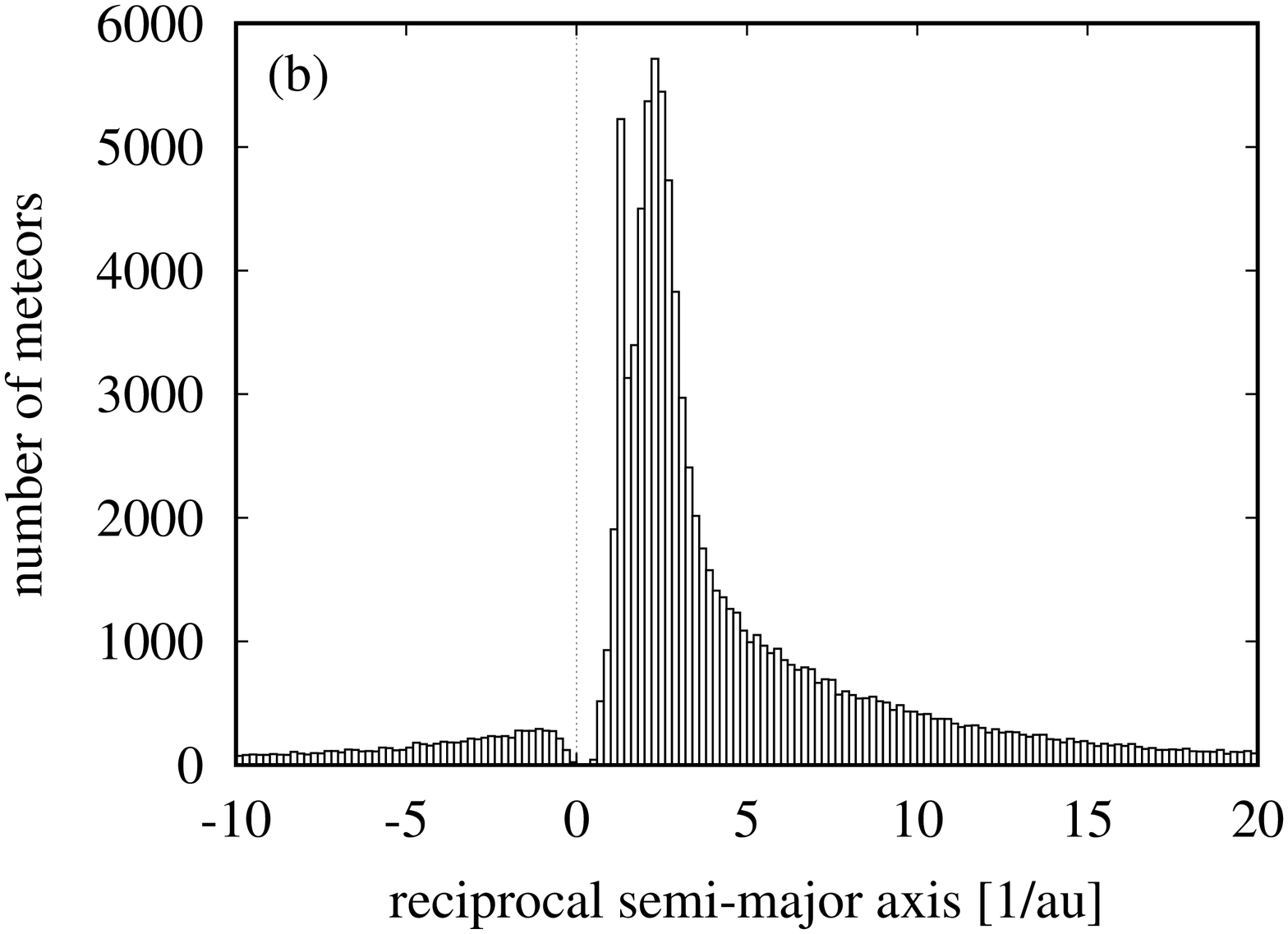}}
\centerline{\includegraphics[width = 5 cm, angle = -90]{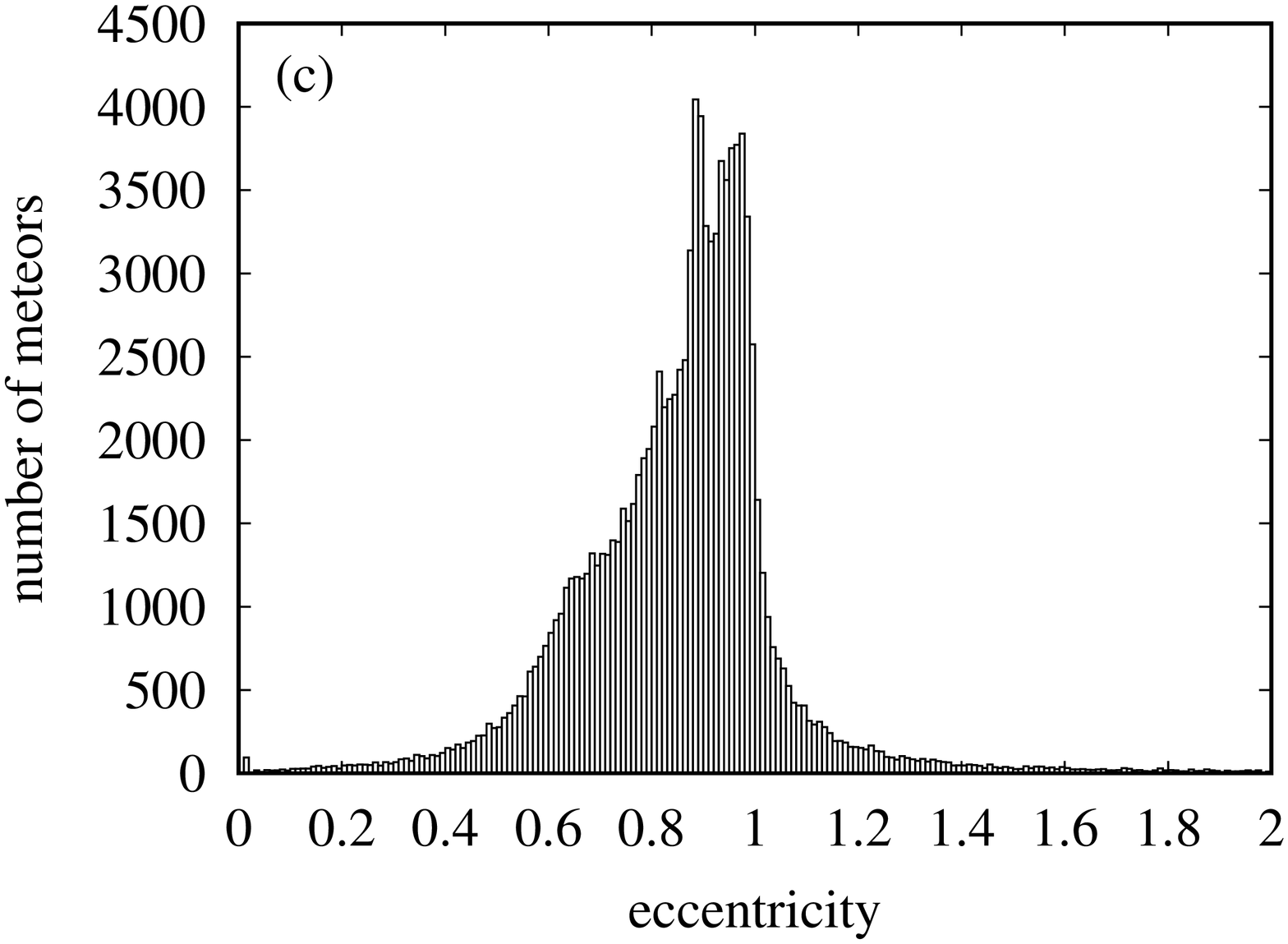}
  \includegraphics[width = 5 cm, angle = -90]{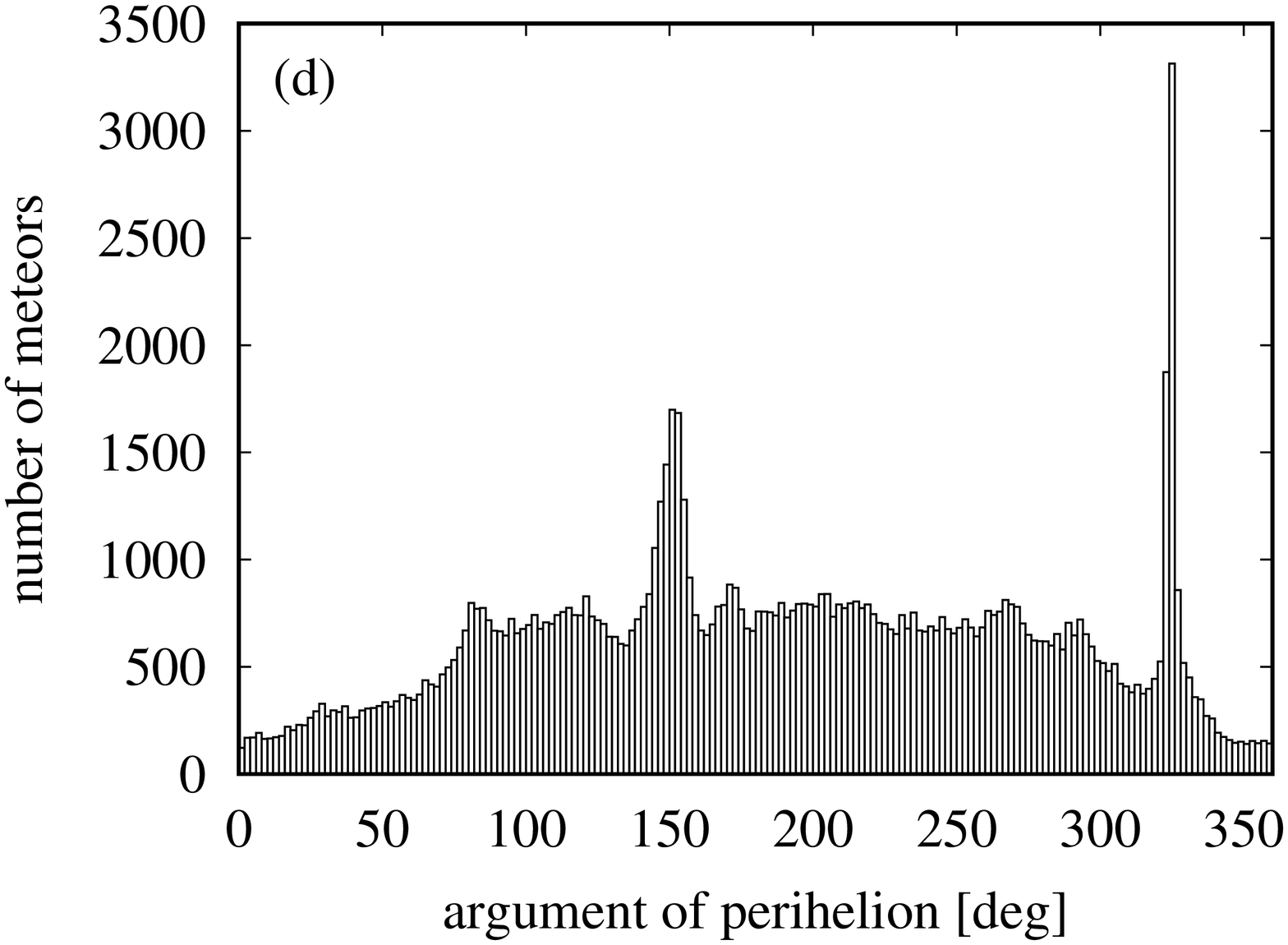}}
\centerline{\includegraphics[width = 5 cm, angle = -90]{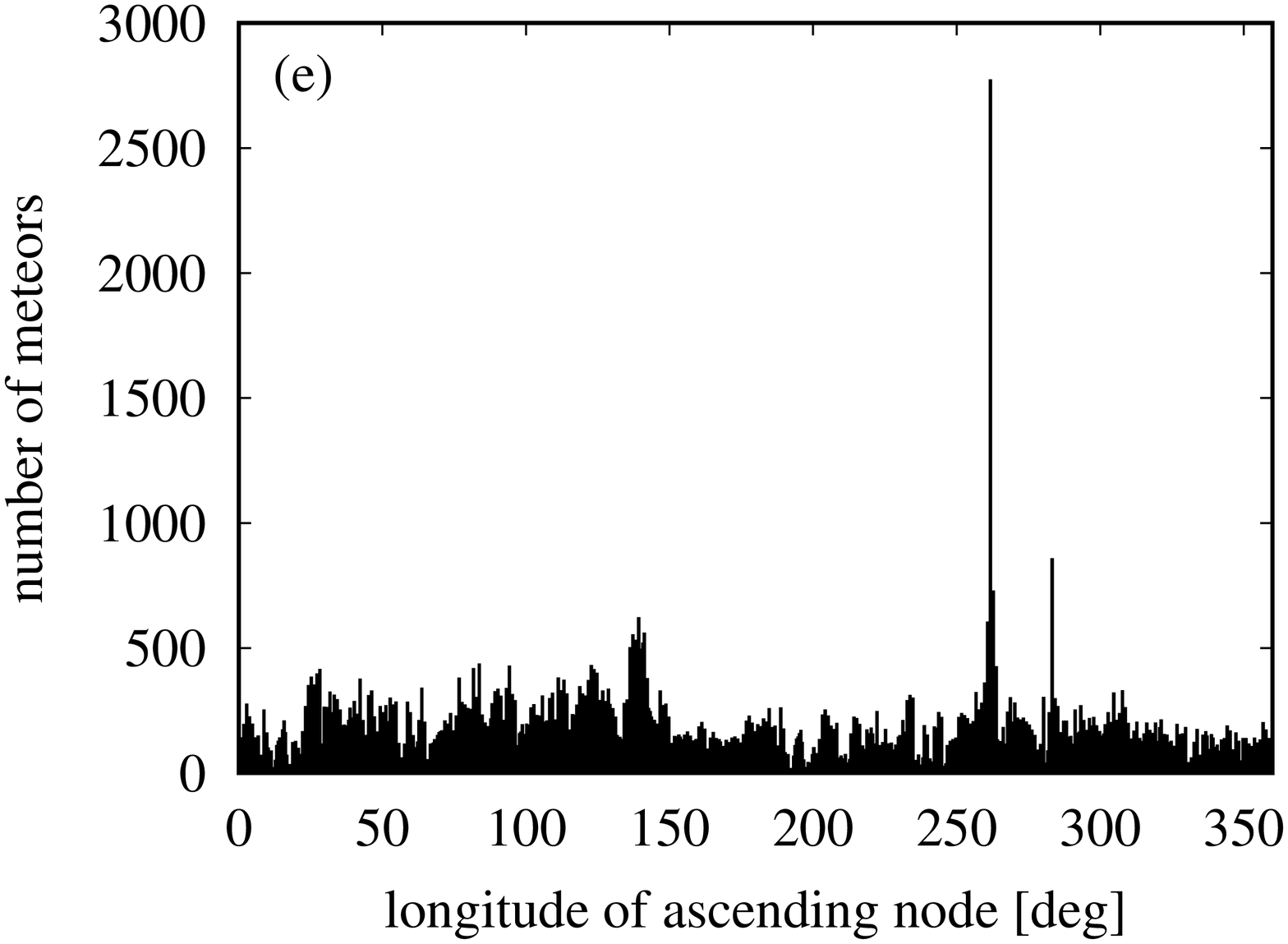}
  \includegraphics[width = 5 cm, angle = -90]{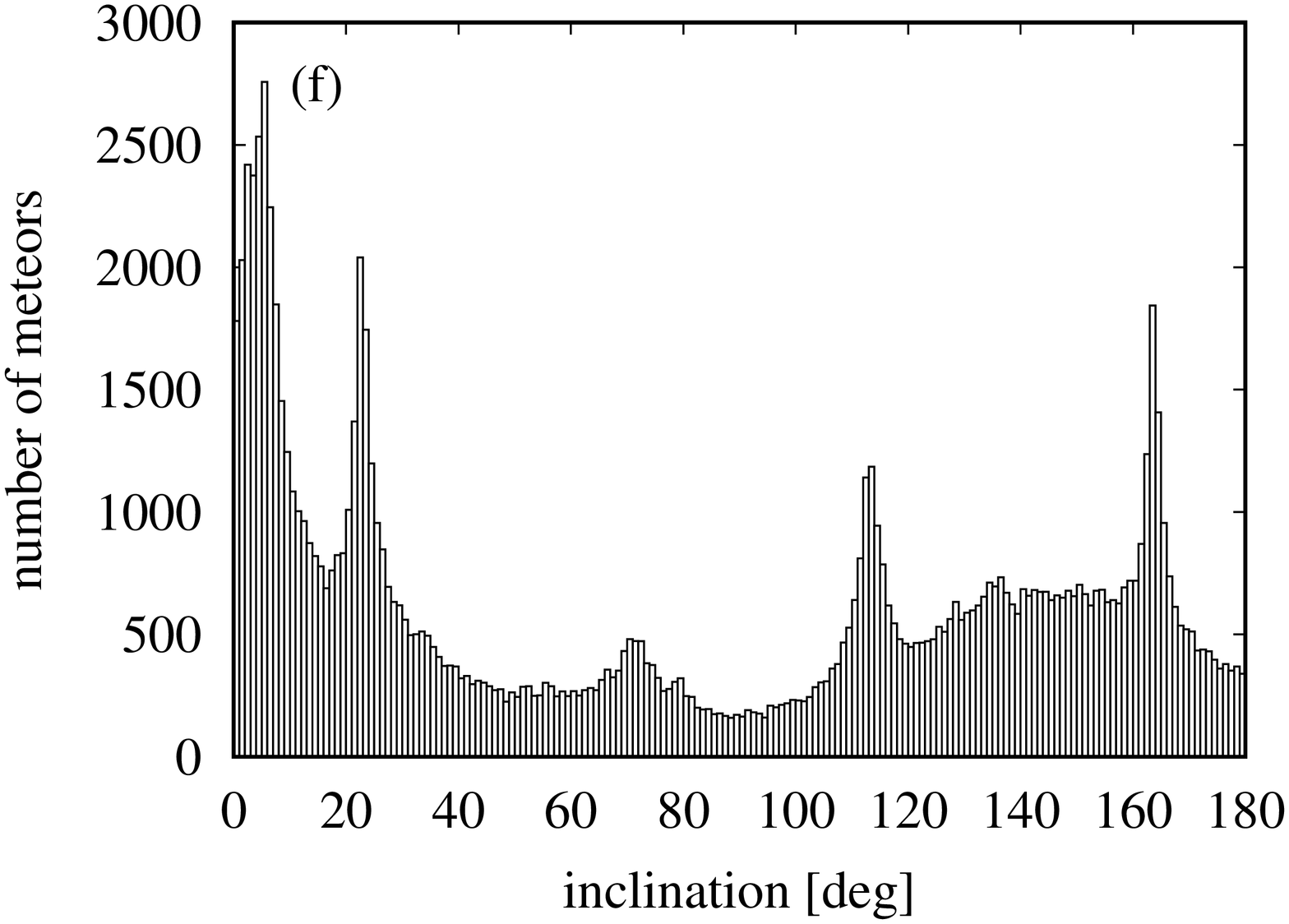}}
\caption{The distributions of perihelion distance (panel a), reciprocal semi-major axis (b), eccentricity (c), argument of perihelion (d), longitude of ascending node (e), and inclination (f) of the video meteors in the CAMS MDC database.}
\label{fig:eledistr}
\end{figure*}

In the current 
OD, each new sample of the orbits submitted is saved separately (though in a consistent form) and can also be downloaded separately. This rule has already been applied to all of the video orbits in the database. The orbital database consists of more than 450 000 orbits, see Table \ref{tab:orbits}. The magnitude distributions of the meteors caused by the meteoroids with known orbital data are shown in Figure \ref{fig:histF}, separately for the meteors detected by photographic (a) and video (b) techniques (photographic meteors and video meteors, hereafter). In Figure \ref{fig:eledistr}, there are, furthermore, the orbital elements of the meteoroids observed by video technique. Only the Cameras for Allsky Meteor Surveillance (CAMS, \cite{2011Icar..216...40J,2016Icar..266..355J,2016Icar..266..384J,2016Icar..266..371J}) video meteors were included in the database at the time of plotting the graphs. 
\begin{table}[!ht]
\centering
\begin{tabular}{lr}
\hline
 {Type of list} &   {Number of showers}   \\
\hline
{List of All Showers} & 917 \\
{List of Established Showers} & 112 \\
{Working List} &  781   \\
{List of Shower Groups} &  24 \\
{List of Removed Showers} &  44 \\
\hline
\end{tabular}
\label{tab:showers}
\caption{IAU MDC shower lists in October 2021.\\ }
\end{table}

\begin{figure*}
\centerline{\includegraphics[width = 15 cm, angle = 0]{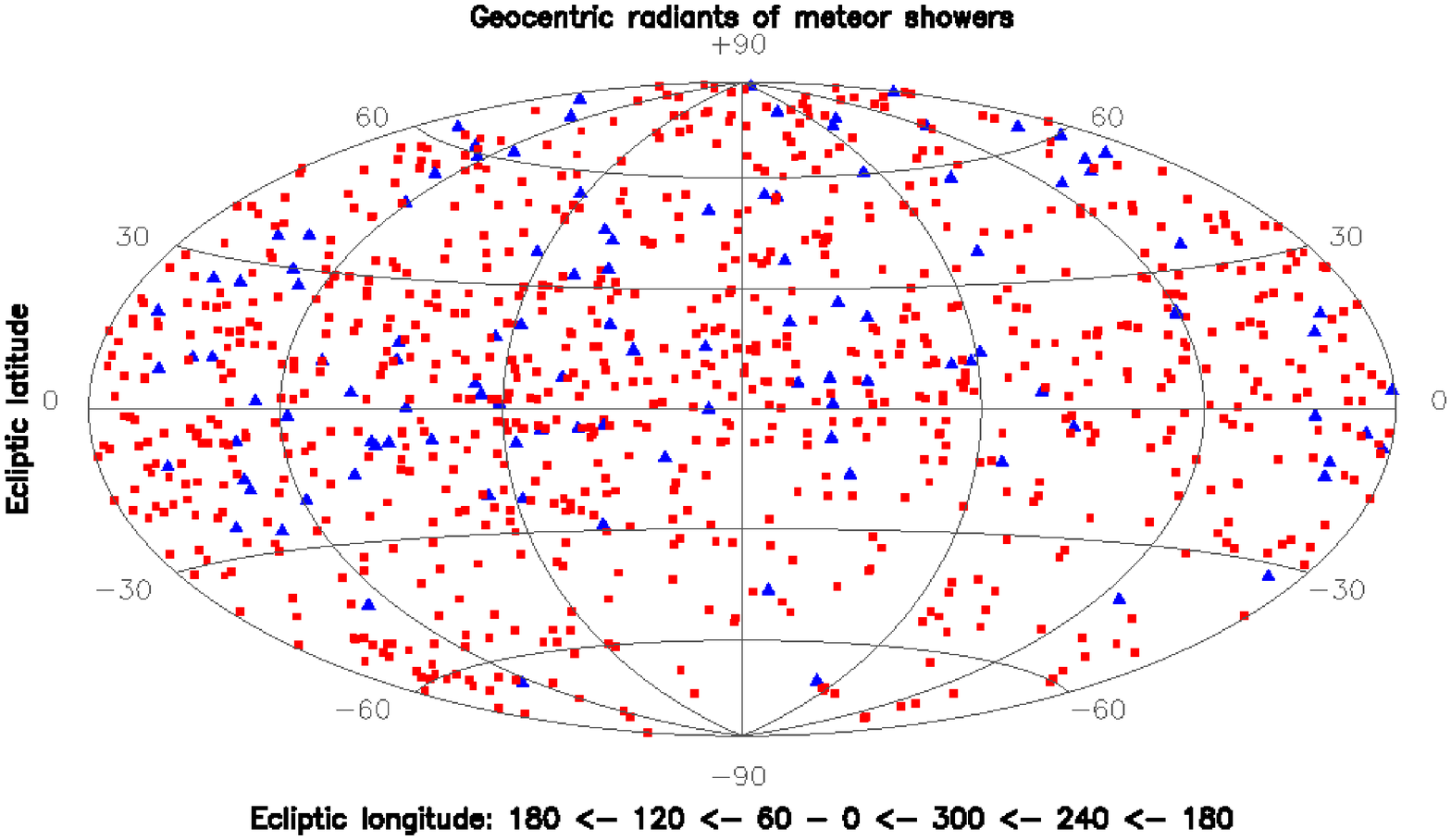}}
\caption{Ecliptical coordinates of the mean radiants of the meteor showers in Aitoff-Hammer's projection. $781$ radiants from the MDC working list are marked with red squares. $112$ shower radiants from the Established list are marked with blue triangles. }
\label{fig:aitoffhammer}
\end{figure*}

Currently, the SD component contains over 900 meteor showers; their names are grouped in the form of five lists, see Table \ref{tab:showers}. The first list consists of the names of all unique showers actually registered in the database. 450 of them are represented by more than one set of parameters (so called solutions). The names of the showers already approved by the IAU are also listed separately in the List of Established Showers. 
The Working List constitutes showers that have yet to be confirmed (by other authors or other data). 
The List of Meteor Shower Groups consists of groups/complexes that have been suggested in a scientific publication. In Figure \ref{fig:aitoffhammer} we plotted the radiants of the meteor showers from the Working and Established lists on the whole celestial sphere. As one can see, in the northern sphere of the ecliptic system, we have slightly more meteor radiants.

Removed showers remain in the MDC in a separate list. The list contains the names of showers that were previously on the Working List and were excluded from it for various, very individual reasons. The problem of removing unnecessary data from the Working List was discussed during the Meteoroids 2019 conference,  (\cite{2020P&SS..18204821J}). 
The Working List may include duplicates and data of very low statistical significance; for example, a shower (steam) identified by means of two orbits only. It was decided that any shower (stream)
would be removed from the Working List if a work recommending such a decision was published.
After a recommendation for removal, the MDC will move the shower to the List of Removed Showers and add a note giving the reason for the removal on the MDC Web site. Any removed shower can eventually return to the Working List after such a recommendation has been published.
However, it should be noted that an incomplete record (e.g. no orbital data) is not a reason for removal, as long as the proposed shower is uniquely identified by the geocentric parameters.

\section{How to submit new data}
The current requirements for submission and the points of contact are on the website, listed separately for orbital and shower data.
\subsection{New orbits of meteoroids}
A new set of orbital data sent to the MDC has to be accompanied by a paper or papers, which should have been published in a peer reviewed journal before submission to the MDC. The paper(s) should provide a description of the observational facility used to detect the meteors and a description of the method of processing the observations, as well as the way of calculating the parameters presented. References to this paper(s) must be sent to the IAU MDC
together with the data (it is assumed that the papers will be cited by users). If the authors send another set of data gained using the same observational equipment and the same method of data processing, the accompanying paper is not required.

For each meteor, the complete set of compulsory parameters must be given. The compulsory parameters are:
\begin{enumerate}
  \item date of the meteor detection (in the form: year, month, day and fraction of day, giving the time of meteor detection in U.T.),
  \item right ascension and declination of the geocentric radiant (in degrees),
  \item geocentric and heliocentric velocities (in km$\,$s$^{-1}$),
  \item perihelion distance of meteor orbit (in AU),
  \item its numerical eccentricity,
  \item argument of perihelion (in degrees),
  \item longitude of ascending node (in degrees), and
  \item inclination (in degrees).
\end{enumerate}

The angular parameters should be referred to the actual equinox of J2000.0. Ideally, at least some of these parameters should be given with their error limits.
   
Besides the compulsory parameters, the authors of new data on the meteors can also supply some of 22 additional parameters for every meteor. The list of 21 of those parameters currently accepted by the IAU MDC was published in a paper by \cite{2020P&SS..19205008N}, Table 1.
The 22nd parameter, duration of meteor, was added recently, in the process of incorporating the SonotaCo data (\cite{2009JIMO...37...55S,2016JIMO...44...42S,SonotaCo_etal2021}).

Before the proper sending of new orbital data, the authors are requested to communicate the way of sending and format of the data with the MDC-OD team\footnote{mdc\underline{~}orbits@ta3.sk}.

\subsection{Delivery of meteor showers}
Observers may send data for both unknown and well-known meteoroid streams to the SD. 
Data provided of a stream which already exist in the MDC will be used to confirm its real existence. Modern methods of identifying meteoroid streams do not guarantee that the results obtained were not obtained by chance.

The shower database is not an archive of all meteoroid information. It only stores those parameters that are necessary to determine whether we are dealing with a new stream or with a well-known stream. 
Hence the geocentric longitude of the Sun, the radiant coordinates, and geocentric velocity are required, as well as  the heliocentric elements of the orbit. However, only the geocentric parameters are obligatory, while the orbital elements are very much appreciated. 
The basic planes of the reference system used in the MDC are the mean equator and the mean ecliptic of the epoch J2000. The moment of time of the shower activity is represented by the ecliptic longitude of the Sun. In brief, the data sent to the MDC may include: 
\begin{enumerate}
    \itemsep=.1mm
    \item Activity – the year of observed shower activity. If observed regularly - the word {\it annual} should be given.
    \item  S. Lon – the mean solar ecliptic longitude at the moment of observation of 
    the members of the shower. 
    (degrees, epoch J2000).
    \item RA – mean geocentric right ascension of the shower radiant (degrees, epoch J2000).
    \item DE – mean geocentric declination of the shower radiant (degrees, epoch J2000).
    \item dRA – radiant drift in right ascension (not obligatory, degrees RA per degree of solar longitude, epoch J2000).
    \item dDE – radiant drift in declination (not obligatory, degrees DE per degree of solar longitude, epoch J2000).
    \item VG – mean geocentric speed (km$\,$s$^{-1}$).
    \item a – mean semi-major axis (AU).
    \item q – mean perihelion distance (AU).
    \item e – mean eccentricity.
    \item Peri – mean argument of perihelion (degrees, epoch J2000).
    \item Node – mean longitude of ascending node (degrees, epoch J2000).
    \item Incl – mean inclination of the orbital plane (degrees, epoch J2000).
    \item N – number of meteors used to determine the mean radiant and orbit.
    \item OT – code of the technique used for the meteor shower observation: P – photo, R – radar, T – TV, video, C- CCD, V – visual.
    \item Information where the submitted meteor data will be published. 
    \item Also, the author may propose unique name of the submitted shower, as well as its unique 3-letter code.
\end{enumerate}
All meteoroid data should be sent to the MDC as an ASCII file, according to the template available on the MDC website.   
Next, the MDC verifies the correctness of the proposed stream name and 3 letter code. If necessary, the MDC determines the correct name and code. From this moment, within 6 months, using the obtained shower name and code, the author should publish a paper describing his/her submission. Publications in any peer-reviewed scientific journal are accepted, but also in amateur journals such as WGN or MeteorNews.

As the mean values of the meteoroid parameters, most authors give the arithmetic means of the corresponding parameters of the stream members. Other authors provide median values of the individual parameters. However, both such approaches do not guarantee consistency between the given parameters, e.g. the mean value of the semi-major axis is usually not equal to the semi-axis calculated using arithmetic means or medians of eccentricity and perihelion distance. This means that the mean values of the orbital elements reported in the MDC do not represent the orbit of a meteoroid stream in terms of celestial mechanics. 

On the MDC website, there are templates of two files: the shower mean data
\footnote{\url{https://www.ta3.sk/IAUC22DB/MDC2007/Etc/streamMDC-Template.txt}}
and the Look-up table
\footnote{\url{https://www.ta3.sk/IAUC22DB/MDC2007/Etc/streamLT-Template.csv}},
which the user must complete and submit to the MDC. 
The first file contains mean parameters describing the submitted shower and which are included in the kernel of the shower database.  
The second file, the so-called 'Look-up table', contains the data of individual meteoroids, the members of a given stream. These are members of the stream, the data of which were used to determine the average values of the stream parameters. The content of the Look-up table was established at the Meteoroids 2019 conference, (see \cite{2020P&SS..18204821J}) during the business meeting of the Working Group on Meteor Shower Nomenclature (IAU Commission F1: Meteors, Meteorites, and Interplanetary Dust). 

Each Look-up table must contain the following information for each meteor on which the new
identification is based:
\begin{enumerate}
    \itemsep=.1mm
    \item CurNum – current number of the meteor in the Look-up table.
    \item SolLon – ecliptic longitude of Sun at the meteor instant (degrees, J2000).
    \item SCELoG – Sun centered ecliptic longitude of the geocentric radiant (degrees).
    \item ELaG – ecliptic latitude of the geocentric radiant (degrees, J2000).
    \item VG – geocentric velocity [km$\,$s$^{-1}$].
    \item IAUNo – IAU numerical code of the shower.
    \item IAUCod – IAU 3 letter code of the shower, (not obligatory).
    \item CatCod – code of the source catalogue of the meteor, (not obligatory).
    \item MetCod – meteor code given in the source catalogue, (not obligatory).
\end{enumerate}
More details on the required data format of the Look-up table records are given in  the template on the MDC website.

The Look-up tables provided to the MDC will allow a more complete insight into the
meteoroid streams submitted to the database. They contain information about shower duration,
as well as radiant and speed dispersion. The MDC user, by comparing the contents of the tables, will therefore be able to assess whether an identified 'new' stream is already in the MDC.

\section{Conclusions}

As a result of the changes in the mode of operation of the MDC shower base, more than 40 streams were moved to the List of Removed Showers or added as another solution of a previously known stream. These were mostly duplicates. We are convinced that the introduced changes, a critical assessment of the database content and the way it functions, 
have led to an improvement in the quality of data contained in the SD, and make the lists more valuable for users.

It will be recalled that before publishing, each new meteoroids stream must receive a unique name, the IAU numeric and 3-letter code from the MDC. Moreover, each new entry to the MDC must be published in a scientific journal or meteor amateur journal like WGN (the Journal of the IMO) or MeteorNews. 
In order to avoid permanent deletion from the MDC, the published manuscript describing the study must be sent to the MDC within half a year of requesting the shower names and codes. Each new submission to the MDC should be accompanied by the corresponding ''Lookup table'' that gives the shower members' parameters.

Concerning the OD component of the MDC database, its video part will be upgraded in a near future; specifically, version 2 of the CAMS data will be replaced by version 3. Altogether, the new version of the CAMS data contains 471577 video meteors.

The MDC database is not perfect. It still contains incorrect data: mistakes and errors, inconsistent values of the parameters, missing information about the data source etc. Furthermore, the functionality of the database could be improved. Hence, we appreciate every critical remark related to the content and the operation of the MDC.

\section*{Acknowledgments}
The authors like to acknowledge: Reiner Arlt, David Asher, Rhiannon Blaauw, Steve Hutcheon, Sirko Molau, Ned Smith, Francisco Oca\~{n}a, \v{Z}eljko Andrei\'{c}, Damir \v{S}egon, Denis Vida, Allan Mulof, Paul Roggemans, Galina Ryabova, Masahiro Koseki, Alfredo Dal'Ava J\'{u}nior, Mikiya Sato, Diego Janches, Ji\v{r}\'{i} Borovi\v{c}ka, Zuzana Ka\v{n}uchov\'{a}, Ivan Sergey and Peter Jenniskens who in recent years helped improve the IAU MDC contents.

This research has made use of NASA's Astrophysics Data System Bibliographic Services.

\bibliographystyle{unsrt}
\bibliography{IMC2021_proceeding_Arxiv} 
\end{document}